\documentclass{aastex63}

\received{\today}
\submitjournal{Research Notes of the AAS}

\shorttitle{GBT Calibration}
\shortauthors{Goddy et al.}

\graphicspath{{./}{figures/}}

\begin{document}

\title{L-band Calibration of the Green Bank Telescope from 2016-2019}

\correspondingauthor{Julian Goddy}
\email{jgoddy@haverford.edu}

\author[0000-0002-8933-9574]{Julian Goddy}
\affiliation{Haverford College \\
370 Lancaster Avenue \\
Haverford, PA, 19041, USA}

\author[0000-0002-3746-2853]{David V. Stark}
\affiliation{Haverford College \\
370 Lancaster Avenue \\
Haverford, PA, 19041, USA}

\author[0000-0003-0846-9578]{Karen L. Masters}
\affiliation{Haverford College \\
370 Lancaster Avenue \\
Haverford, PA, 19041, USA}


\section{Introduction} \label{sec:intro}

The Robert C. Byrd Green Bank Telescope (GBT) is being used to conduct the HI-MaNGA survey \citep{Masters2019}, a program to measure 21cm emission from galaxies in the Mapping Nearby Galaxies with Apache Point Observatory (MaNGA) survey \citep{Bundy2015} which began in 2016. Accurate flux calibration is important to this and all astronomical observing.  The default GBT calibration at L-band (1.4GHz/20cm) comes from the use of a noise-diode of measured antenna temperature ($T_A$). After an observation, the noise diode is turned on, allowing an immediate calculation of the telescope's gain in units of $T_A/{\rm counts}$. One can convert flux densities from $T_A$ to Jy, $S$, using
\begin{equation}
S=\frac{2kT_A}{A\eta},
\end{equation}
where $A$ is the telescope collecting area, $\eta$ is the telescope aperture efficiency, and $k$ is the Boltzman constant.  A complete calibration requires knowing both the antenna temperature of the noise diode, $T_{\rm cal}$, and $\eta$. $T_{\rm cal}$ has been measured in a laboratory to an accuracy of ${\sim}15-20\%$\footnote{http://www.gb.nrao.edu/GBTCAL/}. However, the effective strength of noise diodes can vary over time, and the tabulated values of $T_{\rm cal}$ have not been updated since 2005.  

Using standard astronomical calibrator sources, we have monitored the flux calibration of the GBT in L-band from 2016--2019 using both continuum scans and position-switched (PS) spectroscopic observations. Both types of observations show that the GBT calibration at L-band is remarkably stable over a period of four years, but the GBTIDL routines underestimate fluxes by $\sim20\%$.

\newpage
\section{Methods and Results}

As part of our observing program for HI-MaNGA \citep{Masters2019}, the GBT pointing solution was checked regularly from 2016-2019 using observations known as ``Autopeaks", where the telescope scans across a bright point source. Our Autopeak scans were conducted at 1.4 GHz using the Digital Continuum Receiver (DCR). The following calibration sources were used over our observing program: 3C48, 3C123, 3C147, 3C161, 3C218, 3C227, 3C249\_1, 3C286, 3C295, 3C309\_1, 3C348, and 3C353. The maximum amplitude of an Autopeak scan is the flux density of the source in uncalibrated counts. We then used the $T_{\rm cal}$ to calibration to estimate the physical flux density of the target, to compare to published values\citep{Ott1994}.

In 2019, we additionally conducted PS spectroscopic observations of calibrator sources using the Versatile GBT Astronomical Spectrometer (VEGAS). These observations match the observing setup and ``ON-OFF" pattern of  HI-MaNGA science observations (with the exception that scans are one minute instead of five minutes) and thus allow a consistency check that any observed calibration offset does not depend on observing mode. As with the Autopeak observations, we calculated the flux density of observed sources using the standard $T_{\rm cal}$ values to compare to published values \citep{Ott1994}.

\begin{figure}[bht]
\centering
\includegraphics[angle=0,scale=0.31]{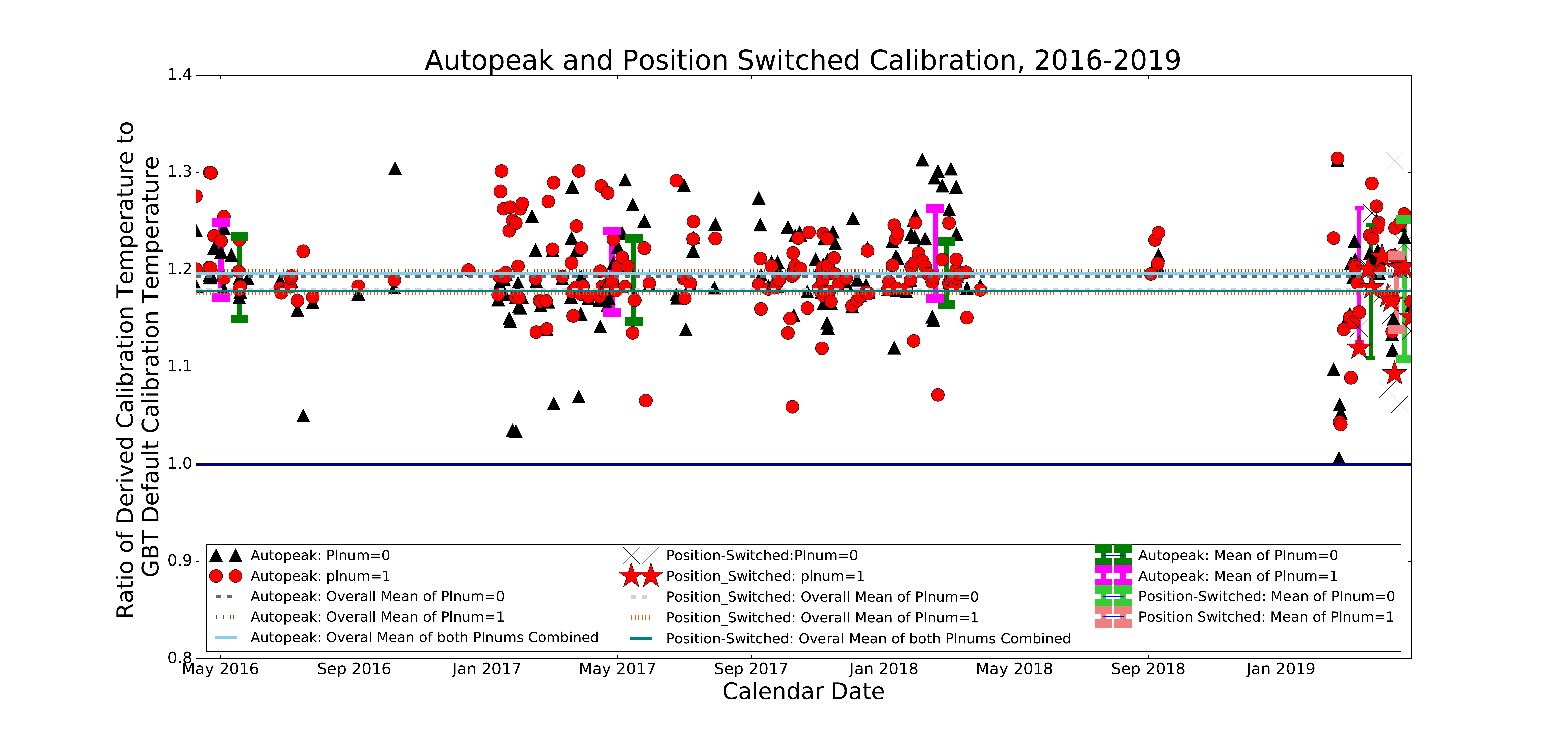}
\caption{Ratio of the true flux to the measured flux using the default $T_{\rm cal}$ as a function of time for both Autopeak and PS observations.  Black triangles and red circles show the two polarizations for the Autopeak data; black crosses and red stars show the two polarizations for PS data. The overall mean for both polarizations individually and combined is shown using horizontal line, with means and standard deviations for each year shown with error bars.}
\label{figure:plots}
\end{figure}

For both the Autopeak and PS observations, we define a ratio $r$ such that: 
\begin{equation}
    r = \frac{{\rm Published\,Flux}}{\rm Measured\,Flux\,(Default\,T_{\rm cal})}
\end{equation}
In Figure \ref{figure:plots} we plot the measured value of $r$ for each session, as well as the mean and standard deviation (ignoring strong outliers) for the entire time range and each year individually. Across a baseline of four years (and for both polarizations), the overall discrepancy between the true and measured fluxes from Autopeak data is $r = 1.20\pm0.05$ and from PS observations is $r=1.18\pm0.07$. 

Our findings show that the GBT noise diode $T_{\rm cal}$ values have been stable over a period of four years, is offset by 20\% from the value implied by observations of standard astronomical flux calibrators. This 20\% offset is within the stated error on the calibration routines, but is a systematic error which can be corrected for. Although the GBT's aperture efficiency could also contribute to the offset in flux calibration, it is not expected to change measurably over time (D. Frayer, private communication), so we expect the offset is due to the $T_{\rm cal}$ value which is used. This analysis highlights the need to regularly check the default GBT calibration using observations of standard astronomical calibrators.

\acknowledgments
We gratefully acknowledge the Haverford College KINSC office, as well as the National Science Foundation's support of the Keck Northeast Astronomy Consortium's REU program through grant AST-1005024. The majority of this work was done as part of a  summer project funded by the Haverford Provost Office. The Green Bank Observatory is a facility of the National Science Foundation operated under cooperative agreement by Associated Universities, Inc.

\end{document}